\documentclass[epsfig]{article}
\usepackage{graphicx}
\usepackage{amsmath}

\input epsf.sty
\newtheorem{theorem}{Theorem}
\newtheorem{acknowledgement}[theorem]{Acknowledgement}

\input{tcilatex}
\begin{document}

\title{\textbf{Ground State Energy of Extended Hubbard Model by
Self-Consistent RPA }}
\author{S.Harir$^{1},$M.Bennai$^{1,3\thanks{%
Corresponding autor: bennai\_idrissi@yahoo.fr,m.bennai@univh2m.ac.ma}}$and
Y.Boughaleb$^{1,2\thanks{%
Member of Hassan-II Academy of Sciences and Tachnology}}$ \\
%EndAName
$^{1}${\small \ Laboratoire de Physiques de la Mati\`{e}re Condens\'{e}e. }\\
{\small Facult\'{e} des Sciences Ben M'Sik, Universit\'{e} Hassan
II-Mohammedia Casablanca,\ Maroc. }\\
$^{2}${\small \ LPMC, Facult\'{e} des Sciences d' Eljadida, Universit\'{e}
Chouaib Doukkali, Maroc.}\\
$^{3}${\small \ Groupement Nationale de Physique des Hautes Energies, Focal
point, LabUFR-PHE, Rabat, Morocco.}}
\date{}
\maketitle

\begin{abstract}
Using the SCRPA, we study the intersite interaction effect on the dynamics
of N electrons system. We have considered an extended Hubbard model
including intrasite and intersite interactions, and we have applied this
model to a system of two neighbouring atoms containing a free electron. The
application of SCRPA to this model allows us to study the intersite
interaction effect on the ground state and the excitation energies of
system. We show that the repulsive interaction between the electrons of the
neighbouring atoms is the origin of \ an supplementary conductivity of the
system.\bigskip


\textbf{Pacs numbers:} 71.10, -W.75.10.Jm, 72.15.Nj
\end{abstract}

\newpage \newpage

\section{Introduction}

The discovery of High-Temperature superconductivity has motivated a
considerable effort in physics of strongly correlated electronic systems,
and many theoretical models have been proposed\cite{HSUP}. The Hubbard model 
\cite{Hub,Beenen} is one of the simplest and more general description of an
interacting electrons system on a lattice. In its simplest form, it includ
the usual kinetic energy of electrons and the competition between the
on-site electron-electron interaction. The standard Hubbard model is defined
by the second quantized Hamiltonian \cite{Hub,Beenen}:

\begin{equation}
H=\sum_{i\neq j,\sigma }t_{ij}.c_{i,\sigma }^{^{\dagger }}.c_{j,\sigma
}+U.\sum_{i}n_{i,\uparrow }.n_{i,\downarrow }
\end{equation}
The first term of the eqs(1) represents the kinetic energy of electrons, and
each electron has a possibility of hopping between different lattice sites. $%
c_{j,\sigma }$ is the annihilation operator of the electron at a lattice
site $j$ with spin index $\sigma $. $c_{i,\sigma }^{^{\dagger }}$is the
creation operator of the electron at a lattice site $i$, so \ $t_{ij}$ is
the hopping integral from the site $j$ to the site $i$. The second term
represents the intrasite coloumb interaction with energy $U$, where $%
n_{i,\sigma }$ is the number operator of electrons at the site $i$ with spin 
$\sigma $ .

Recently, the Random Phase Approximation (RPA)\cite{RPA}, was used to
resolve the standard Hubbard model\cite{Jemai,Rabhi}. The RPA is an approach
which treats seriously the correlations of system, and attempt to minimise
the system energy. The standard Hamiltonian Hubbard must be developed as
function of the creation and annihilation operators of the pair
particle-hole(p-h), because our RPA regroups the electrons system on pair:
particle-hole. The application of RPA to standard Hubbard gives non linear
coupled equations, where the resolution is done by a SCRPA\cite{Dukelsky,Vd}%
. In this paper, we consider an extended Hubbard model\cite{EHM}, where the
intersite coulomb interactions are introduced. This model was shown to
describe many interesting properties of high TC superconductors materials 
\cite{EHMSUP}. In this work, we apply the SCRPA method to this extended
Hubbard model, and study the intersite interaction effect on the ground
state and excitation energy of system. We show that the repulsive
interaction between the electrons of the neighbouring atoms is the origin
of\ a supplementary conductivity of the system.\bigskip

\section{Extended Hubbard Model}

The standard Hubbard model with intrasite interaction explains some
important physical phenomena like High-Temperature superconductivity \cite%
{Beenen}, Mott-transition\cite{Zhang}. To explain other physical phenomena
observed in different areas of the solid state physics like magnetic and
transport properties, it is convenient to take into account also the
intersite interaction resulting from original coloumb repulsion modified by
the polaronic effect. The extended Hubbard Hamiltonian is then given by\cite%
{Calegari,Grabiec}:

\begin{equation}
H=\sum_{i\neq j,\sigma }t_{ij}.c_{i,\sigma }^{^{\dagger }}.c_{j,\sigma
}+U.\sum_{i}n_{i,\uparrow }.n_{i,\downarrow }+\frac{1}{2}\sum_{i\neq
j,\sigma }V_{ij}^{(1)}.n_{i,\sigma }.n_{j,\sigma }+\frac{1}{2}\sum_{i\neq
j,\sigma }V_{ij}^{(2)}.n_{i,\sigma }.n_{j,-\sigma }
\end{equation}
U denotes the effective intrasite coloumb interaction. $V_{ij}^{(1)}$( $%
V_{ij}^{(2)}$) describes the effective intersite coloumb interaction between
the electrons in the lattice sites i and j, with the same spins (opposite
spins). $V_{ij}^{(1)}$ and $\ V_{ij}^{(2)}$ are not necessary equal. The
model(2) cannot be solved in a general case. There is however, a special but
non trivial case of finite number of sites, which possesses exact analytical
solution\cite{Jemai}.

In this work we will limit ourselves to a simple case, and will apply the
general formalism of SCRPA to the two sites problem. We consider a closed
chain in one dimension, with periodic boundary conditions $N=2$. Our
physical system is then equivalent to two neighbouring atoms containing a
free electron. The Hamiltonian of the system is: 
\begin{eqnarray}
H_{II} &=&-t\sum_{\sigma }.(c_{1,\sigma }^{^{\dagger }}.c_{2,\sigma
}+c_{2,\sigma }^{^{\dagger }}.c_{1,\sigma })+U.(n_{1,\uparrow
}.n_{1,\downarrow }+n_{2,\uparrow }.n_{2,\downarrow }) \\
&&+V_{1}.\sum_{\sigma }n_{1,\sigma }.n_{2,\sigma }+V_{2}.\sum_{\sigma
}n_{1,\sigma }.n_{2,-\sigma }  \notag
\end{eqnarray}
where $t=-t_{12}=-t_{21}$.

In order to apply the approximation SCRPA to the Hubbard model, it is
necessary, first, to apply the Hartree-Fock approximation(HF) to the Hubbard
model. In the HF method, we write the Hamiltonian(3) as function of
quasiparticles operators, wich allow us to obtain the excitation spectrum of
independent quasiparticles. The states $\left\vert HF\right\rangle $ are
defined as: $\left\vert HF\right\rangle =a_{k_{i},\uparrow }^{^{\dagger
}}.a_{k_{i},\downarrow }^{^{\dagger }}.\left\vert vac\right\rangle $, where $%
a_{k,\sigma }^{^{\dagger }}$ is the annihilation operator of the mode $%
(k,\sigma )$; $a_{k;\sigma }$ is related to $c_{j;\sigma }$with the usual
Fourier transformation:

\begin{equation}
c_{j,\sigma }=\frac{1}{\sqrt{N}}.\sum_{k,\sigma }a_{k,\sigma }.\exp (-i.%
\overrightarrow{k}.\overrightarrow{R_{j}})
\end{equation}
k is the momentum of state $\left\vert k,\sigma \right\rangle $. The
periodic boundary conditions suppose that $c_{N\text{ }+\text{ }j,\sigma
}=c_{j,\sigma }$. With this condition, eqs(4) gives $\exp (-i.%
\overrightarrow{k}.\overrightarrow{R_{j}})=1$, which have two solutions in
the first Brillouin zone: $k_{1}=0$ and $k_{2}=-\pi $. Thus the Hamiltonian
is then written as: 
\begin{equation}
H_{HF}=E_{HF}+\sum_{\sigma }\left\{ \varepsilon _{1}.n_{1,\sigma
}+\varepsilon _{2}.n_{2,\sigma }\right\}
\end{equation}
This expression shows that in the HF approximation, the physical system has
two possible states $\left\vert HF\right\rangle $ and $\left\vert
HF\right\rangle ^{\ast }$. $\left\vert HF\right\rangle $ ($\left\vert
HF\right\rangle ^{\ast }$ ) is the Hartree-Fock ground state (excited state
), with the momentum: $k_{1}=0;$ below (and $k_{2}=-\pi ;$ above ) the Fermi
momentum, and the excitation energy: $\varepsilon _{1}$ (and $\varepsilon
_{2}$), where $\left\vert HF\right\rangle =a_{k_{1},\uparrow
}^{+}.a_{k_{1},\downarrow }^{+}.\left\vert vac\right\rangle $ and $%
\left\vert HF\right\rangle ^{\ast }=a_{k_{2},\uparrow
}^{+}.a_{k_{2},\downarrow }^{+}.\left\vert vac\right\rangle $. As in ref 
\cite{Grabiec}, we define the HF quasiparticle operators by: $b_{1,\sigma
}=a_{k_{1},\sigma }$ and\ $b_{2,\sigma }=a_{k_{2},\sigma }$ . We have then $%
b_{k,\sigma }\left\vert HF\right\rangle =0$, for all k.

In normal ordering of $b_{1,\sigma }$ and $b_{2,\sigma }$, the Hamiltonian$%
(3)$ becomes:

\begin{equation}
H=H_{HF}+H_{k=0}+H_{k=-\pi }
\end{equation}
where

\begin{equation*}
H_{k=0}=\frac{U+V_{2}}{2}\left( \tilde{n}_{k_{2},\uparrow }-\tilde{n}%
_{k_{1},\uparrow }\right) \left( \tilde{n}_{k_{2},\downarrow }-\tilde{n}%
_{k_{1},\downarrow }\right) +\frac{V_{1}}{4}.\sum_{\sigma }\left( \tilde{n}%
_{k_{2},\sigma }-\tilde{n}_{k_{1},\sigma }\right) ^{2}
\end{equation*}
\begin{equation*}
H_{k=-\pi }=-\frac{U-V_{2}}{2}\left( J_{\uparrow }^{-}+J_{\uparrow
}^{^{\dagger }})(J_{\downarrow }^{-}+J_{\downarrow }^{^{\dagger }}\right) -%
\frac{V_{1}}{4}.\sum_{\sigma }\left( J_{\sigma }^{-}+J_{\sigma
}^{^{^{\dagger }}}\right) ^{2}
\end{equation*}
With $\ $

\begin{equation*}
J_{\sigma }^{-}=b_{1,\sigma }\,b_{2,\sigma },~J_{\sigma }^{^{\dagger
}}=\left( J_{\sigma }^{-}\right) ^{^{\dagger }},~\tilde{n}_{k_{i},\sigma
}=b_{i,\sigma }^{\dagger }\,b_{i,\sigma }
\end{equation*}
$H_{k=0}$ and $H_{k=-\pi }$ take into account the correlation between the
number operators of the type: $\tilde{n}_{k_{i},\sigma }\tilde{n}%
_{k_{j},\sigma ^{\prime }}$ in the ground state: $k_{1}=0$ (below the Fermi
momentum ) and between the magnetic momentum operators of the type: $%
J_{\sigma }^{^{\dagger }}.J_{\sigma ^{\prime }}^{-}$ in the excited state: $%
k_{2}=-\pi $ (above the Fermi momentum).$~$\newpage

\section{\protect\bigskip Self Consistent Random Phase Approximation}

\subsection{Formalism}

In order to apply the Formalism of SCRPA to the Hubbard model, it is
convenient to use the particle-hole(ph-RPA) approximation, which regroup the
physical system on pair. We can then define the RPA excitation operator as:

\begin{equation}
Q_{v}^{^{\dagger }}=\sum_{p,h}(x_{ph}^{v}.b_{p}^{^{\dagger
}}.b_{h}^{^{\dagger }}-y_{ph}^{v}.b_{h}.b_{p})
\end{equation}
Where $h$ (and $p$) are the momentum below (and above) the Fermi momentum.
Eqs(7) shows that the excitation in the ph-RPA is done only by the creation
or (annihilation) of pair: particle-hole via the operator $b_{p}^{^{\dagger
}}.b_{h}^{^{\dagger }}$ ($b_{h}.b_{p}$) with the amplitude $x_{ph}^{v}$ ($%
y_{ph}^{v}$). The corresponding excited state of this excitation operator is 
$\left\vert v\right\rangle =Q_{v}^{^{\dagger }}.\left\vert RPA\right\rangle $%
, and the corresponding excitation energy is:

\begin{equation}
E_{v}=\frac{\left\langle RPA\right| \left[ Q_{v},\left[ H,Q_{v}^{^{\dagger }}%
\right] \right] \left| RPA\right\rangle }{\left\langle RPA\right| \left[
Q_{v},Q_{v}^{^{\dagger }}\right] \left| RPA\right\rangle }
\end{equation}
Where $\left| RPA\right\rangle $ is the vacuum of this RPA excitation
operator: $Q_{v}\left| RPA\right\rangle =0$

The minimization of $E_{v}$ leads to usual RPA equations of type:

\begin{equation*}
\left( 
\begin{array}{cc}
A & B \\ 
-B^{\ast } & -A^{\ast }%
\end{array}
\right) \left( 
\begin{array}{c}
x^{v} \\ 
y^{v}%
\end{array}
\right) =E_{v}.\left( 
\begin{array}{c}
x^{v} \\ 
y^{v}%
\end{array}
\right)
\end{equation*}

With the relations of the orthonormality conditions of the set $\left\{
Q_{v};Q_{v}^{^{\dagger }}\right\} $, we can express the elements of $A$ and $%
B$ by the RPA-amplitudes, and therefore we have a completely closed system
of equations for amplitudes $x$ and $y$. For our problem, we consider only
the excitation operators, which conserve the spin, where the excitation is
done only by the creation or annihilation of the pair: particle-hole with
the same spin.

\begin{equation}
Q_{v}^{^{\dagger }}=x_{\uparrow }^{v}.k_{\uparrow }^{^{\dagger
}}+x_{\downarrow }^{v}.k_{\downarrow }^{^{\dagger }}-y_{\uparrow
}^{v}.k_{\uparrow }^{-}-y_{\downarrow }^{v}.k_{\downarrow }^{-}
\end{equation}

With $K_{\sigma }^{^{\dagger }}=b_{2,\sigma }^{^{\dagger }}.b_{1,\sigma
}^{^{\dagger }}/\sqrt{1-\left\langle M_{\sigma }\right\rangle }$, $\
K_{\sigma }^{-}=b_{1,\sigma }.b_{2,\sigma }/\sqrt{1-\left\langle M_{\sigma
}\right\rangle }$ and $M_{\sigma }=\widehat{n}_{1,\sigma }+\widehat{n}%
_{2,\sigma }$, where the mean values $\left\langle ...\right\rangle $ are
taken with respect to the RPA vacuum $(Q_{v}\left| RPA\right\rangle =0)$.
The SCRPA equation can then be written in the form:

\begin{equation}
\left( 
\begin{array}{cccc}
A_{\uparrow \uparrow } & A_{\uparrow \downarrow } & B^{\uparrow \uparrow } & 
B_{\uparrow \downarrow } \\ 
A_{\downarrow \uparrow } & A_{\downarrow \downarrow } & B_{\downarrow
\uparrow } & B_{\downarrow \downarrow } \\ 
-B_{\uparrow \uparrow } & -B_{\uparrow \downarrow } & -A_{\uparrow \uparrow }
& -A_{\uparrow \downarrow } \\ 
-B_{\downarrow \uparrow } & -B_{\downarrow \downarrow } & -A_{\downarrow
\uparrow } & -A_{\downarrow \downarrow }%
\end{array}
\right) .\left( 
\begin{array}{c}
x_{\uparrow }^{v} \\ 
x_{\downarrow }^{v} \\ 
y_{\uparrow }^{v} \\ 
y_{\downarrow }^{v}%
\end{array}
\right) =E_{v}.\left( 
\begin{array}{c}
x_{\uparrow }^{v} \\ 
x_{\downarrow }^{v} \\ 
y_{\uparrow }^{v} \\ 
y_{\downarrow }^{v}%
\end{array}
\right)
\end{equation}
Where the SCRPA matrix elements are given by:

\begin{equation*}
A_{\sigma \sigma ^{\prime }}=\left\langle \left[ K_{\sigma }^{-},\left[
H,K_{\sigma ^{\prime }}^{^{\dagger }}\right] \right] \right\rangle
\end{equation*}
and

\begin{equation*}
B_{\sigma \sigma ^{\prime }}=\left\langle \left[ K_{\sigma }^{-},\left[
H,K_{\sigma ^{\prime }}^{-}\right] \right] \right\rangle
\end{equation*}
\ \ \qquad The orthonormality relations of the set $\left\{
Q_{v};Q_{v}^{^{\dagger }}\right\} $, give:

\begin{equation*}
\begin{array}{cc}
A_{\uparrow \uparrow }=A_{\downarrow \downarrow }=A & A_{\uparrow \downarrow
}=A_{\downarrow \uparrow }=A^{\prime } \\ 
B_{\uparrow \uparrow }=B_{\downarrow \downarrow }=B & B_{\uparrow \downarrow
}=B_{\downarrow \uparrow }=B^{\prime }%
\end{array}%
\end{equation*}
From the Hamiltonian given in eqs(10), we can writte the SCRPA matrix
elements as: $\ \ A=B+2.t$, \ $A^{\prime }=B^{\prime }$, \ where:

\bigskip 
\begin{eqnarray}
A &=&2.t+\left( U-V_{2}\right) .\sqrt{\frac{1-\left\langle M_{\downarrow
}\right\rangle }{1-\left\langle M_{\uparrow }\right\rangle }}%
.\sum_{v}x_{\uparrow }^{v}(y_{\downarrow }^{v}+x_{\downarrow }^{v}) \\
&&-\frac{V_{1}}{2}.\left( \frac{1}{1-\left\langle M_{\sigma }\right\rangle }%
-\sum_{v}(x_{\sigma }^{v}.x_{\sigma }^{v}+y_{\sigma }^{v}.y_{\sigma
}^{v}+2.x_{\sigma }^{v}.y_{\sigma }^{v})\right) \text{ }
\end{eqnarray}
and 
\begin{equation}
A^{\prime }=\frac{U-V_{2}}{2}.\frac{1}{1-\left\langle M_{\sigma
}\right\rangle }
\end{equation}
where

\begin{equation*}
\left\langle M_{\sigma }\right\rangle =\frac{2\;\sum\limits_{\nu
}\;|y_{\sigma }^{\nu }|^{2}}{1+2\;\sum\limits_{\nu }\;|y_{\sigma }^{\nu
}|^{2}}
\end{equation*}
so, the ph-RPA matrix takes the form:

\begin{equation*}
\left( 
\begin{array}{cccc}
A & A^{\prime } & A-2.t & A^{\prime } \\ 
A^{\prime } & A & A^{\prime } & A-2.t \\ 
2.t-A & -A^{\prime } & -A & -A^{\prime } \\ 
-A^{\prime } & 2.t-A & -A^{\prime } & -A%
\end{array}
\right) \left( 
\begin{array}{c}
x_{\uparrow }^{v} \\ 
x_{\downarrow }^{v} \\ 
y_{\uparrow }^{v} \\ 
y_{\downarrow }^{v}%
\end{array}
\right) =E_{v}.\left( 
\begin{array}{c}
x_{\uparrow }^{v} \\ 
x_{\downarrow }^{v} \\ 
y_{\uparrow }^{v} \\ 
y_{\downarrow }^{v}%
\end{array}
\right)
\end{equation*}
This ph-RPA matrix has two positive roots:

\begin{equation}
\varepsilon _{1}=2.t.\sqrt{\frac{A-A^{\prime }}{t}-1\text{ }}\text{ \ \ \
and \ \ \ \ }\varepsilon _{2}=2.t.\sqrt{\frac{A+A^{\prime }}{t}-1}
\end{equation}
The corresponding eigenvectors are: $V_{1}=\left[ x_{1}.,-x_{1},y_{1},-y_{1}%
\right] \ $and$\ V_{2}=\left[ x_{2}.,-x_{2},y_{2},-y_{2}\right] $,
respectively. Where

\begin{equation}
x_{1}=-\frac{A-A^{\prime }+\varepsilon _{1}}{A-A^{\prime }-2.t}.y_{1}\text{
\ \ \ \ \ \ ;\ \ \ \ \ \ \ }x_{2}=-\frac{A+A^{\prime }+\varepsilon _{1}}{%
A+A^{\prime }-2.t}.y_{1}
\end{equation}
and

\begin{equation}
y_{1}=-\sqrt{2.\left( \frac{A-A^{\prime }+\varepsilon _{1}}{A-A^{\prime }-2.t%
}\right) ^{2}-2}\text{ \ \ \ \ \ \ ;\ \ \ \ \ \ }y_{2}=-\sqrt{2.\left( \frac{%
A+A^{\prime }+\varepsilon _{2}}{A+A^{\prime }-2.t}\right) ^{2}-2}
\end{equation}
So, like the HF approximation, in ph-RPA, our system have tow excitation
energies $\varepsilon _{1}$and $\varepsilon _{2}$, but they are coupled.
Thus in this work, we solve a system of a coupled equations numerically by
iteration leading to a SCRPQ solution which are quasi identical to the exact
result.\newpage

\subsection{\textbf{Results and discussion}}

To show the effect of intersite interaction on the energy of the system, we
have studied th evolution of the ground state and excited energies in term
of the interaction $V_{1}$ and $V_{2}$.

In figure1 we plot the variation of the ground state energy $%
E_{SCRPA}=\left\langle 0\left\vert H\right\vert 0\right\rangle $ as function
of the two parameters of the intersite interaction $V_{1}$ and $V_{2}.$

\FRAME{ftbpF}{5.8271in}{3.8363in}{0pt}{}{}{fig1.png}{\special{language
"Scientific Word";type "GRAPHIC";maintain-aspect-ratio TRUE;display
"USEDEF";valid_file "F";width 5.8271in;height 3.8363in;depth
0pt;original-width 23.0732in;original-height 15.1532in;cropleft "0";croptop
"1";cropright "1";cropbottom "0";filename '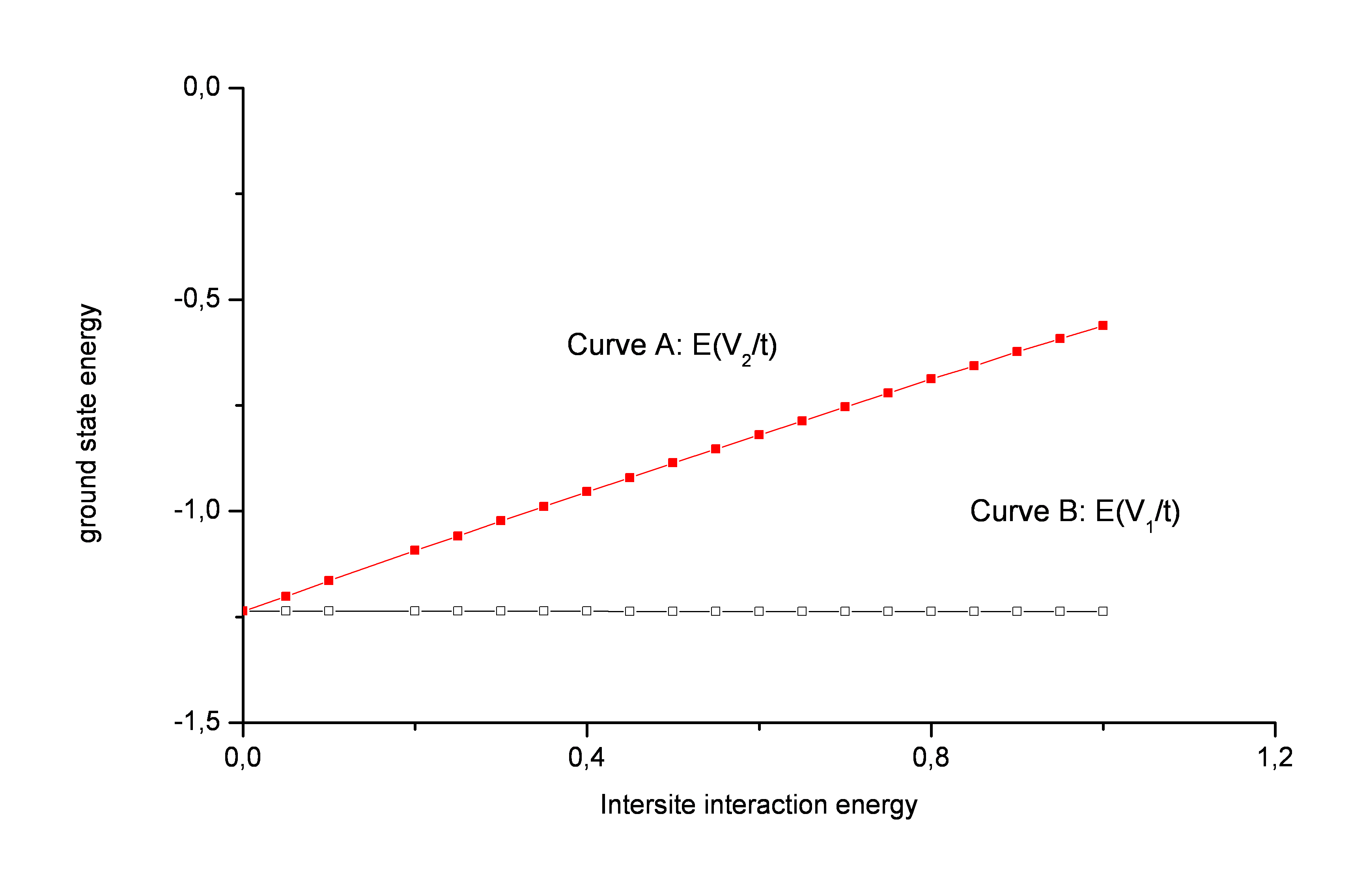';file-properties
"XNPEU";}}

\bigskip\ \ \ \ \ \ \ \ \ \ \ \ \ \ \ \ \ \ \ \ \ \ \ \ \ \ \ \ \ \ \ \ \ \
\ \ \ \ \ \ \ \ \ \ \ \ \ \ \ \ \ \ \ \ \ \ \ \ \ \ \ \ \ \ \ \ \ \ \ \ \ \
\ \ \ \ \ \ \ \ \ \ \ \ \ \ \ \ \ \ \ \ \ \ \ \ \ \ \ \ \ \ \ \ \ \ \ \ \ \
\ \ \ \ \ \ \ \ \ \ \ \ \ \ \ \ \ \ \ \ \ \ \ \ \ \ \ \ \ \ \ \ \ \ \ \ \ \
\ \ \ \ \ \ \ \ \ \ \ \ \ \ \ \ \ \ \ \ \ \ \ \ \ \ \ \ \ \ \ \ \ \ \ \ \ \
\ \ \ \ \ \ \ \ \ \ \ \ \ \ \ \ \ \ \ \ \ \ \ \ \ \ \ \ \ \ \ \ \ \ \ \ \
\qquad \qquad\ \ \ 

The result shows that the ground state energy is quasi independent on $V_{1}$
but, the variation of $E_{SCRPA}$ become more important when we introduce
the intersite interaction with the opposite spins: $V_{2}$. This results can
be explained by the fact that the SCRPA include only, for the fundamental
state, the correlations between the particles with different spins\textbf{:} 
$\left\vert 0\right\rangle =\left( c_{0}^{1}+c_{1}^{1}J_{\uparrow
}^{+}.J_{\downarrow }^{+}\right) \left\vert HF\right\rangle $

Thus the only type of interaction wich is of interest is $V_{2}.$ In the
following we analyse the $V_{2}$ effect on the dynamics of system. In figure
2 and 3 we plot the variation of the gorund state energy and the excitation
energies of SCRPA, respectively, as function of the repulsive intrasite
interaction U for different values of the intersite interaction $V_{2}.$

\FRAME{ftbpF}{5.4129in}{3.8372in}{0in}{}{}{fig2.png}{\special{language
"Scientific Word";type "GRAPHIC";maintain-aspect-ratio TRUE;display
"USEDEF";valid_file "F";width 5.4129in;height 3.8372in;depth
0in;original-width 22.8864in;original-height 16.1867in;cropleft "0";croptop
"1";cropright "1";cropbottom "0";filename '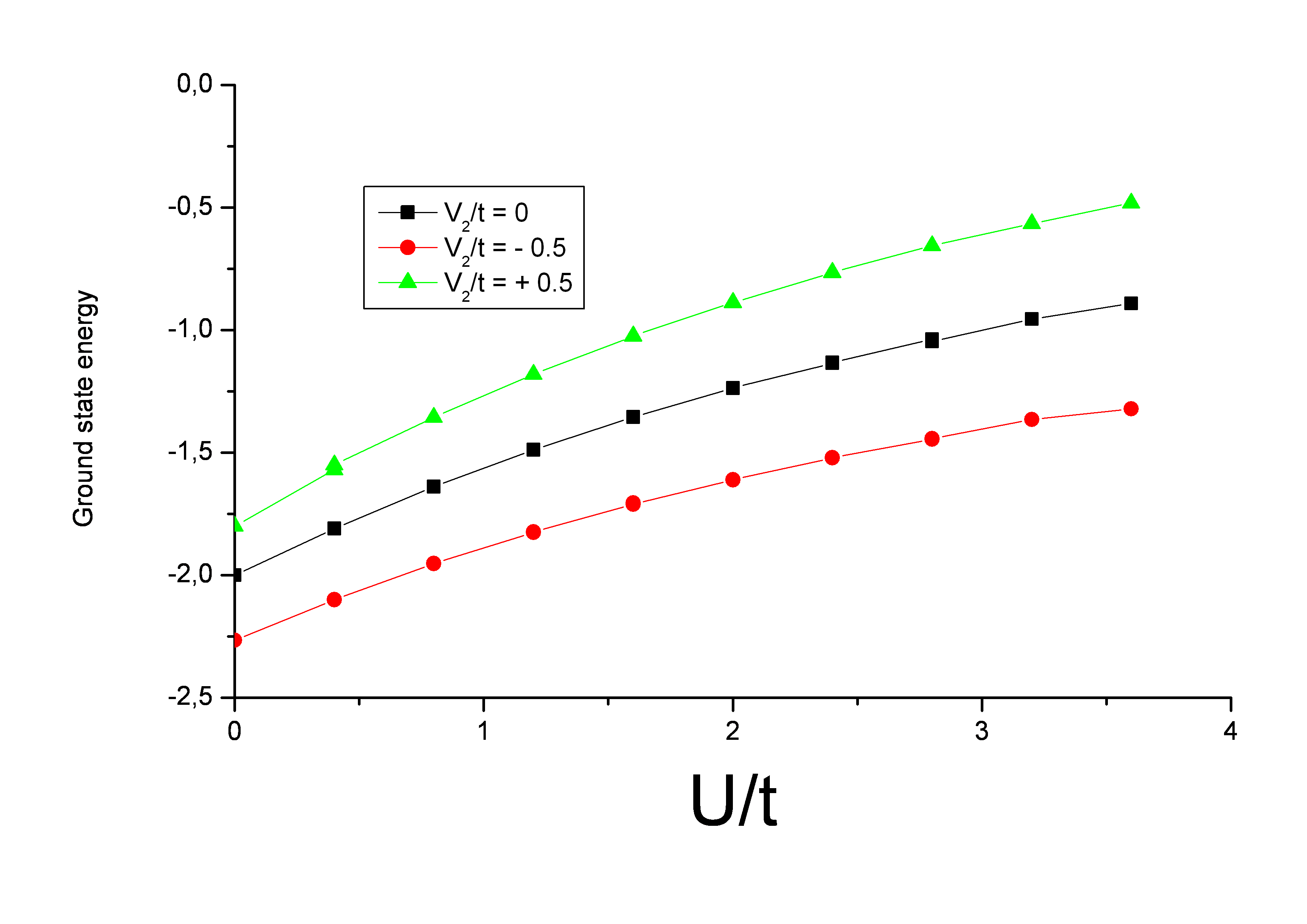';file-properties
"XNPEU";}}

\FRAME{ftbpF}{5.4423in}{3.8363in}{0in}{}{}{fig3.png}{\special{language
"Scientific Word";type "GRAPHIC";maintain-aspect-ratio TRUE;display
"USEDEF";valid_file "F";width 5.4423in;height 3.8363in;depth
0in;original-width 22.793in;original-height 16.0328in;cropleft "0";croptop
"1";cropright "1";cropbottom "0";filename '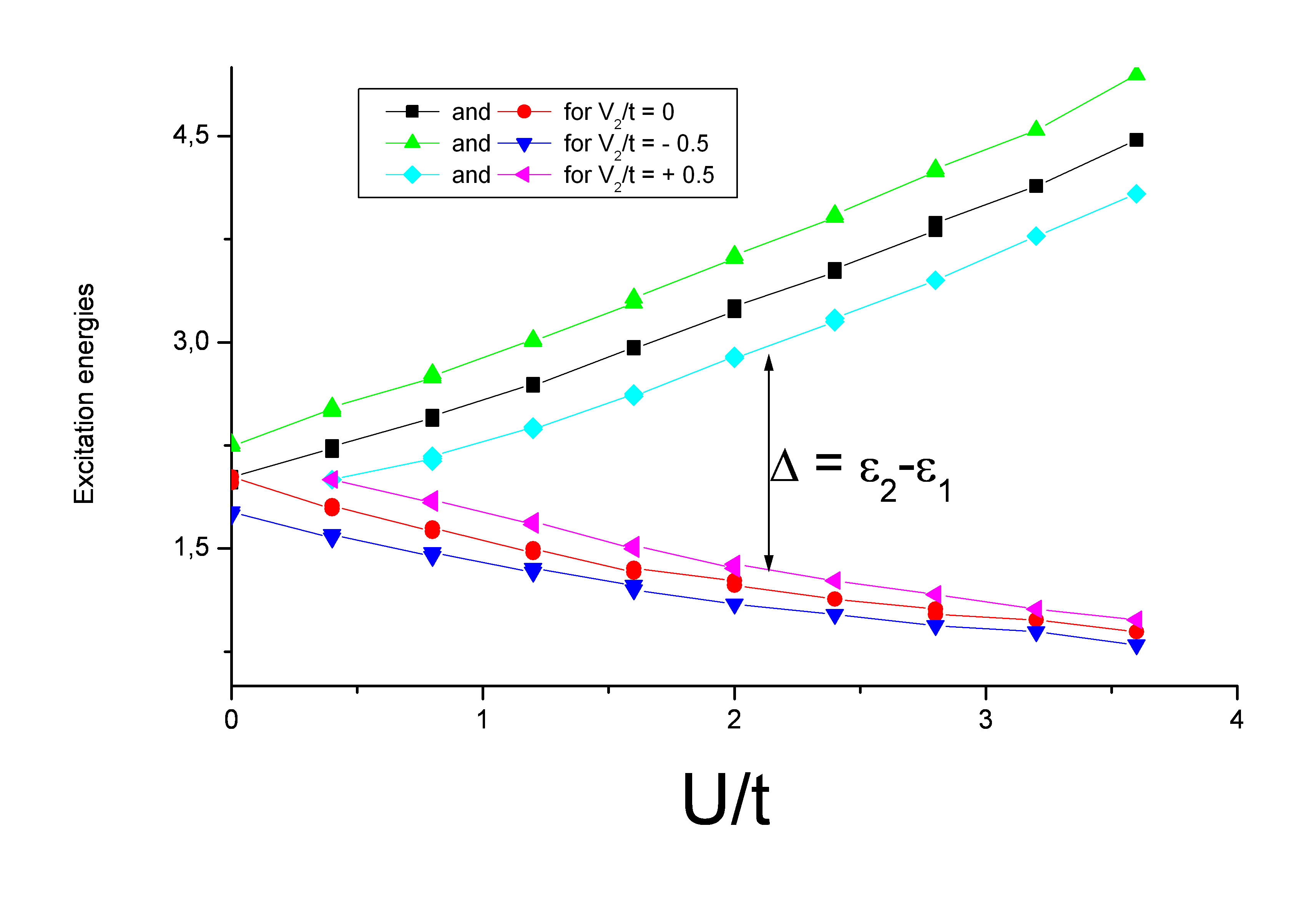';file-properties
"XNPEU";}}

\newpage

The results show that, when $U$ increases, the gap $\Delta =\epsilon
_{2}-\epsilon _{1}$ between the two excitation energies increases too, and
so, the jump of electrons between these two states becomes more difficult.
Thus, we can conclude, that for a fixed value of U, while the intersite
interaction is repulsive (attractive), the gap $\Delta $ becomes less (more)
important. These remarks allow us to assume that repulsive interaction
between the electrons of the neighbouring atoms is the origin of
supplementary conductivity of the system.

\section{\protect\bigskip Conclusion}

In this paper, the SCRPA approximation was used to solve the extended
Hubbard model given in Eqs(3). The quality of the SCRPA method has been
investigated in a previous work by Jemai\cite{Jemai}, in which he has shown
a remarkable agreement between SCRPA method and excact results for the
standard Hubbard model. In our work, we have extended this technic to study
the intersite interaction effects on the dynamics of the electrons in the
two sites with $\left\langle n_{i,\uparrow }\right\rangle =\left\langle
n_{i,\downarrow }\right\rangle $. We have shown that the gap between the
excitation energies: $\left( \Delta =\epsilon _{2}-\epsilon _{1}\right) $
are correlated with the intersite interaction energy $V_{2}$. This result
allow us to suppose that the repulsive intersite interaction (between the
electrons of the neighbouring atoms) is the origin of a supplementary
conductivity of the system. In future work\cite{HBB}, we propose to solve
the 4-sites case (plaquette), which may be very important for the
explanation of high $T_{c}$ superconductivity, by considering the many
plaquette configurations in 2D.

\begin{acknowledgement}
\bigskip The authors thank Professor Rachid Nassif \ for carefully reading
the manuscript and making many constructive comments.
\end{acknowledgement}

\end{document}